# Research Excellence Milestones of BRIC and N-11 Countries

Nadine Rons[1]

[1] Nadine.Rons@vub.ac.be,
Research Coordination Unit and Centre for R&D Monitoring (ECOOM), Vrije Universiteit Brussel (VUB),
Pleinlaan 2, B-1050 Brussels (Belgium)

**Introduction**

While scientific performance is an important aspect of a stable and healthy economy, measures for it have yet to gain their place in economic country profiles. As useful indicators for this performance dimension, this paper introduces the concept of milestones for research excellence, as points of transition to higher-level contributions at the leading edge of science. The proposed milestones are based on two indicators associated with research excellence, the impact vitality profile and the production of review type publications, both applied to a country's publications in the top journals Nature and Science. The milestones are determined for two distinct groups of emerging market economies: the BRIC countries, which outperformed the relative growth expected at their identification in 2001, and the N-11 or Next Eleven countries, identified in 2005 as potential candidates for a BRIC-like evolution. Results show how these two groups at different economic levels can be clearly distinguished based on the research milestones, indicating a potential utility as parameters in an economic context.

**Research question and methodology**

Bibliometric literature includes many comparative studies of scientific production and citation impact at country level. In the last decade 'emerging economies', driven by actual growth as well as by the opening up of a country, have been a frequent subject. The complex context of interactions between the major sectors of universities, industry, and government is addressed in the Triple Helix model (Etzkowitz & Leydesdorff, 1997). This complex interplay results in a clear relation between 'wealth intensity' and 'citation intensity', in particular for countries with lower GDP per person, as demonstrated by King (2004). Measures capturing the level of a country's contribution to the international research scene can therefore represent an important dimension in its economic profile. A wide range of indicators has been used in country studies, from very basic numbers to sophisticated indicators, such as the presently standard field normalized citation rates that were introduced in a context of national research performance (Braun & Glänzel, 1990; Moed et al., 1995). Many are based on the global set of publications, and may invisibly contain very different situations at excellence level. An approach towards excellence is made in studies focusing on the 'best' publications of a country, e.g. using the h-index (Hirsch, 2005) based on the most highly cited papers. The present paper focuses on excellence by (1) using two specific indicators for excellence that lend themselves to the determination of milestones, indicating when a country enters a phase of higher-level research performance, and (2) applying these to the top journals Nature and Science, both internationally recognized media of highest scientific prestige:

- The impact vitality profile was introduced in a context of individual scientists (Rons & Amez, 2009), with as key aspect a sustained progress in high-level performance measured from citing publications. It was tuned for this application by choosing a moving window of only 3 years for citing publications, in order to focus on stability (not bridge fluctuations). The cited publications are limited to Article, Letter, Note and Review, as the standard document types for scientific contributions in citation analysis. The *'impact vitality milestone'* is defined as the most recent year in the impact vitality profile, in which a country enters a period of continuously increasing impact (i.e. with impact vitality values larger than 1), generated by its Nature and Science publications. This year remains fixed as long as the number of citing publications increases, which may continue as long as conditions stay favourable for growth at world level (global volume of research) and local level (investments in research and openness).

- Review type publications are the basis for a measure of esteem for a country's or institution's top researchers introduced by Lewison (2009) as the percentage of reviews in a publication set. The *'review milestone'* is defined as the year of publication of a country's first review in Nature or Science, indicating that the country's maintained support of research brought it to a level where it includes some of the most esteemed researchers. This year remains fixed, unless one would choose to re-evaluate a country's milestone after a period of strongly reduced scientific visibility, possibly related to economic decline or a closed state.

Both milestones are determined for two distinct groups of economically changing countries: the BRIC countries (O'Neill, 2001), and the N-11 countries (O'Neill et al., 2005). Their capacity to distinguish the two groups is examined, testing the potential utility of the research excellence milestones as parameters in an economic context.

**Results and discussion**

All calculations for this paper were made using the on line Web of Science. The impact vitality profiles of most BRIC and N-11 countries are found to proceed from a first phase where growth is repeatedly interrupted by stagnation or decline, to a more stable phase of steady growth, starting at the impact vitality milestone. The exception is China, with continuously growing impact from the beginning. The impact vitality milestone occurs earlier for the BRIC countries than for the N-11 countries. Moreover, as the first phase includes periods of increasing impact of variable length, from 1 up to 11 years with an average of 3 years, milestones for N-11 countries that lie close to the final year of observation are to be regarded as preliminary. Also the review milestone occurs earlier for the BRIC countries than for the N-11 countries, where it has not been reached yet in 4 out of the 11 cases. The two milestones are positively correlated (N=11; r=0,75; p=0,004), not surprisingly as they both stand for a transition to higher-level research performance.

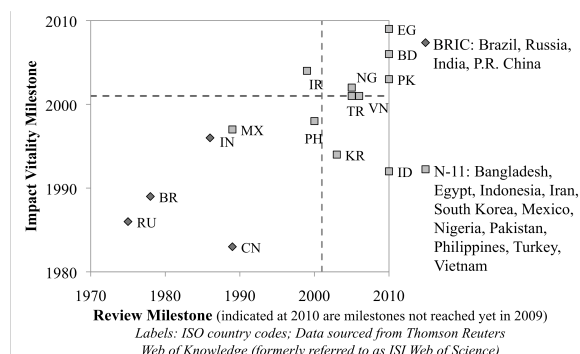

**Figure 1. Research Excellence Milestones of BRIC and N-11 Countries.**

Figure 1 shows both milestones for the BRIC and N-11 countries. Both groups of countries can be clearly distinguished. In 2001, when the BRIC countries were defined (dashed lines), these had both milestones well behind them, unlike the N-11 countries. The only N-11 country approaching the BRIC countries is Mexico, which was pointed out as perhaps the only one having the capacity to become as important globally as the BRIC countries (O'Neill et al., 2005). When compared to the economic parameters for 2005 used in the same paper, both milestones are correlated with GDP, modestly for the review milestone (N=11; r=-0,53; p=0,05) and strongly for the impact vitality milestone (N=15; r=-0,81; p=0,0001), and not significantly correlated (|r|<0,27, p>0,17) with population size and GDP per capita. These observations suggest that neither sheer population size, nor average wealth intensity, are determining factors for a country to reach high research performance levels, but rather well directed resources for investment. Similarly, structural and policy settings determine the Growth Environment Score (GES) introduced to rank the BRIC and N-11 countries' capacities to 'catch up' with developed countries. When compared to these GES scores, the research milestones are not significantly correlated (|r|<0,29, p>0,08), suggesting that research level related parameters could bring an extra dimension into such country analyses.

**Conclusion**

In line with earlier evidence relating research performance to economic parameters, the results show that indicators related to research excellence in particular can add a useful dimension to a country's economic profile. The possibility to distinguish the BRIC and N-11 countries based on the proposed research excellence milestones, indicates the potential utility of parameters for high-level research performance in economic analysis regarding growth expectations.